\documentclass[aps,prd,letterpaper,twocolumn,superscriptaddress,floatfix]{revtex4-1}
\usepackage{graphicx,psfrag,amsmath,amssymb,amsfonts,bbm,latexsym,color,dcolumn,bm}
\usepackage[english]{babel}

\begin{document}
	
\title{Bounds to strong gravity from electron-positron annihilations near threshold}

\author{Nathaniel Alden}

\thanks{nalden@uchicago.edu}

\thanks{Current address: Department of Physics, University of Chicago,  Chicago, IL 60637, USA}

\affiliation{\mbox{Department of Physics and Astronomy, Dartmouth College, 6127 Wilder Laboratory, 
		Hanover, NH 03755, USA}}

\author{Roberto Onofrio}

\thanks{onofrior@gmail.com}

\affiliation{\mbox{Department of Physics and Astronomy, Dartmouth College, 6127 Wilder Laboratory, Hanover, NH 03755, USA}}

\begin{abstract}
  We discuss bounds to strong gravity arising from annihilation of electron and positrons and the corresponding enhancement of the cross section for the production of massive particles. 
Two complementary examples are discussed, the case of the $e^+e^- \to W^+ W^-$ processes, 
with maximum mass and short lifetime for the particle pairs in the final state, and the case of 
electron positron annihilation into heavier charged lepton pairs, $e^+ e^- \to \mu^+ \mu^-$ and $e^+ e^- \to \tau^+ \tau^-$. These bounds may be improved by the next generation of electron-positron colliders, and are complementary to bounds arising from direct graviton searches.
  
\end{abstract}

\maketitle

\section{Introduction}

In the attempt to solve the hierarchy problem without necessarily invoking supersymmetric models, a connection between the standard model and gravity at the Fermi scale is conjectured in several approaches now under scrutiny. Conjectures regarding a connection between gravitation and neutrinos have been discussed since several decades (see for instance \cite{Dicke,Brill}), including efforts to consider weak interactions as the short-range counterpart of gravitation  \cite{Hehl1,Hehl2,Gasperini,Sivaram}, or natural extensions of the successful mixing between electromagnetic and weak interactions \cite{Novello,Loskutov}.
This connection can emerge through several mechanisms proposed more recently, such as introducing extra dimensions \cite{Arkani,Antoniadis,Randall}, four-dimensional models with large number of fermions \cite{Dvali,Calmet}, Kerr-Newman black holes with spin-dragging effects \cite{Burinskii1,Burinskii2,Burinskii3}, and macroscopic gravity containing a long-range residual of the Higgs field \cite{Dehnen1,Dehnen2,Consoli1,Consoli2}, among many attempts.  Moreover, strong gravity at the attometer scale - with the Newtonian gravitational constant reaching the strength of the weak interactions - allows for an interpretation of the Yukawa couplings of fundamental fermions in terms of relative balance between quantum vacuum and gravitation  \cite{Onofrio}. A possible mixing between gravity and weak interactions could also give insights on the chiral structure of the latter \cite{Alexander} - a feature of weak interactions not shared by electromagnetic and color interactions - encouraging their embedding into a geometrodynamics program \cite{Wheeler}. 

This multiform theoretical activity has also motivated searches for gravitational signals at high energy colliders, for instance looking to deviations from purely quantum electrodynamics processes such as photon pair production, $e^+e^- \to \gamma \gamma$, starting from TRISTAN collider \cite{Abe}. The search of anomalous signals indicating graviton emission has been unsuccessful so far, even at the highest energies available at the LHC  \cite{Bound1,Bound2,Bound3}. Also, precision experiments looking for deviations from Newtonian gravity at much lower energies, using macroscopic apparata and force measurements, have 
not evidenced positive signal, although they have enormously extended the exclusion region \cite{Murata}. 

\begin{figure*}[t!]
\centering
\includegraphics[width=0.98\columnwidth]{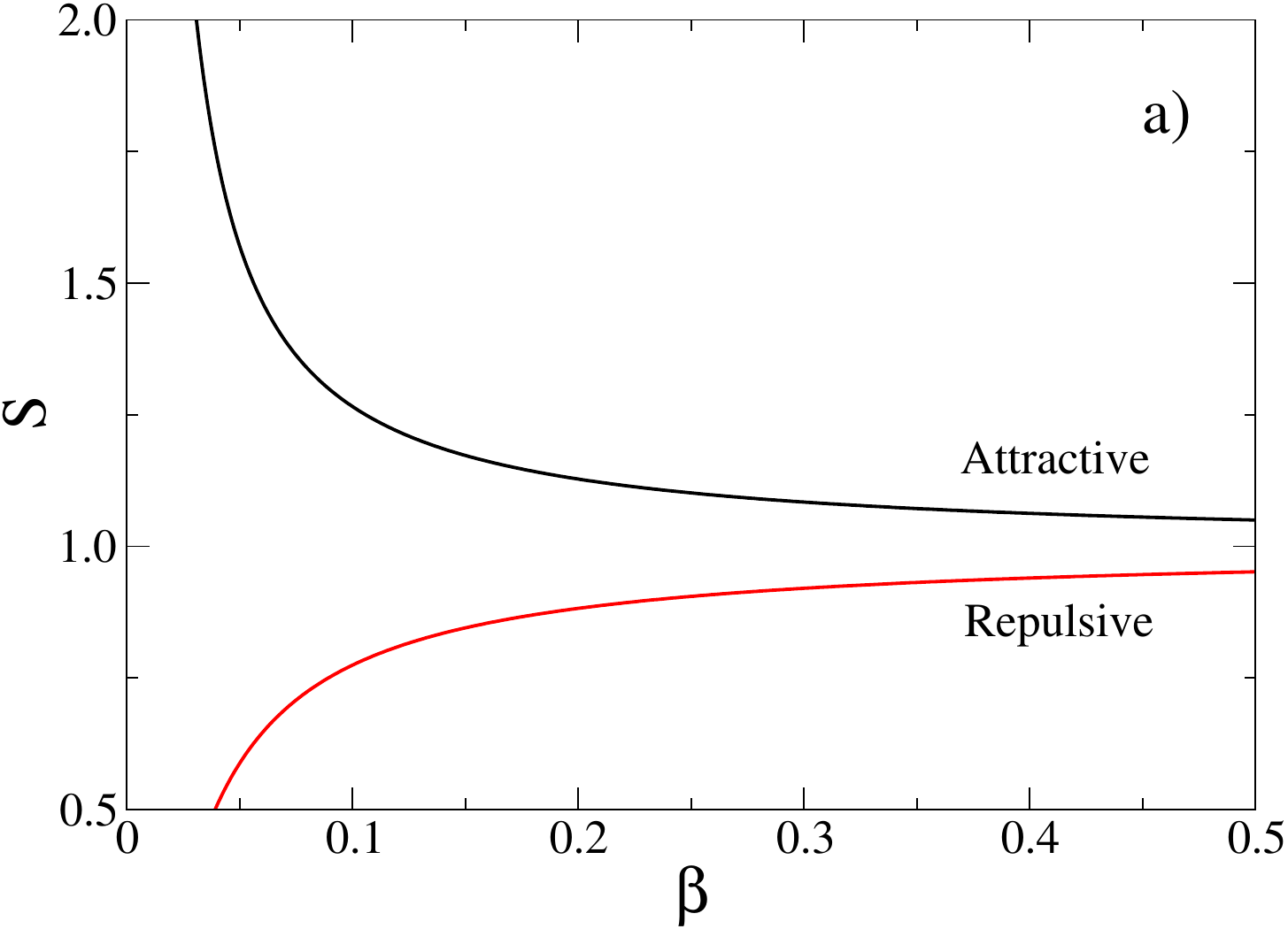}
\includegraphics[width=0.98\columnwidth]{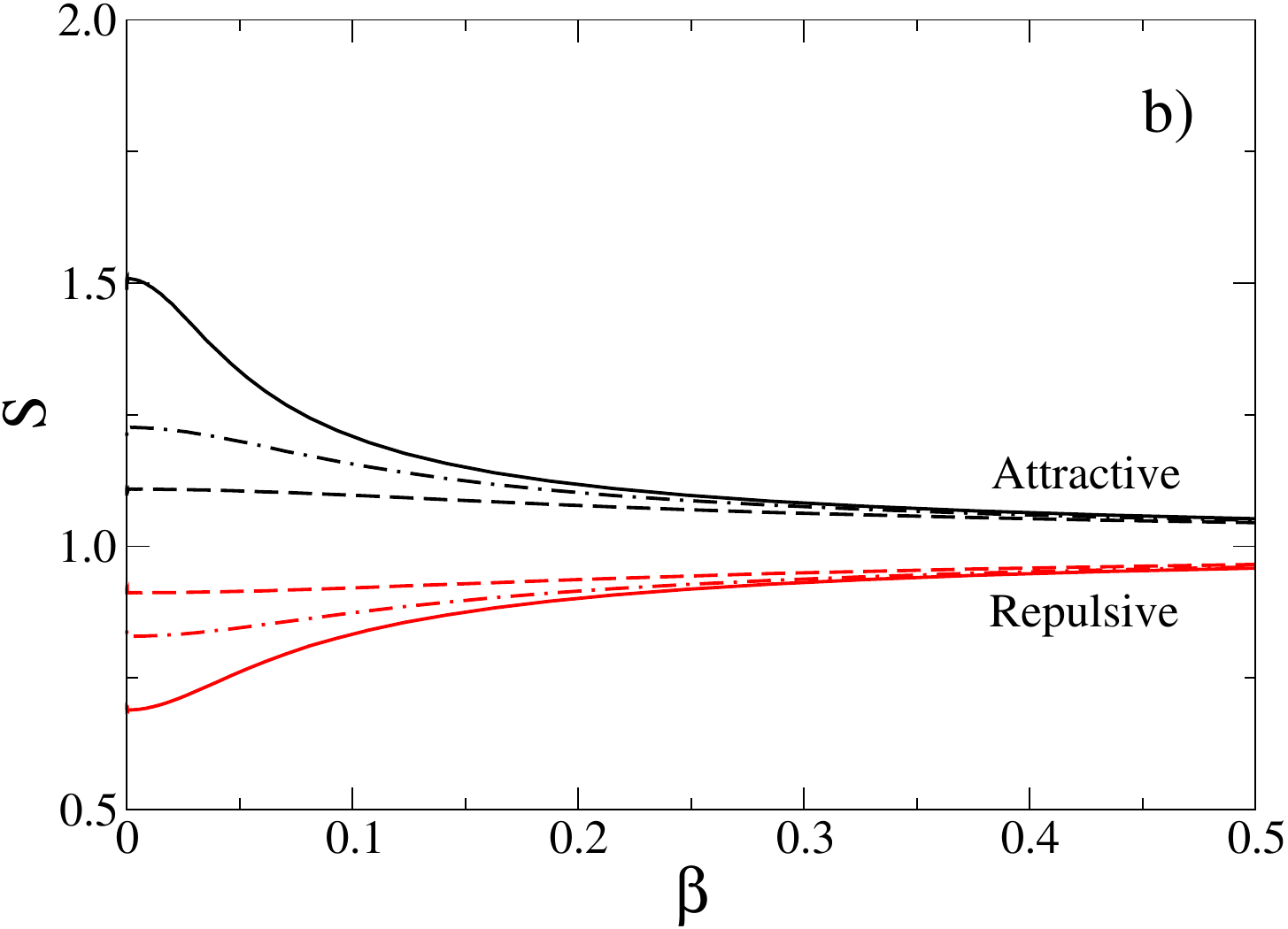}
\caption{Modification of the production cross section due to the presence of a Coulomb-like potential (a) and a Yukawa-like potential (b) versus the factor $\beta=v/c$ up to values for which nonrelativistic approximations as the one in Eq. \ref{Sformula} held. In the Coulomb case the $S$ factor scales as $\beta^{-1}$ at small $\beta$ for attractive interactions, while it goes to zero for repulsive interactions in the same limit. The coupling strength used here for both potentials is $\alpha=1/128$, corresponding to the strength of the electromagnetic interaction at the $Z^0$ pole, as a reference point.  The Yukawa potential is chosen with a mass of the intermediate boson $m = 80~ \mathrm{GeV}$. The pattern is similar to the one of the Coulomb case, except that in this case the $S$ factor saturates below a certain velocity, with a magnitude depending on the particle mass $M$, here represented with the cases of $M$=2 TeV (continous line), $M$=1 TeV (dot-dashed line), and M=0.5 TeV (dashed line), see similar plots in \cite{Lattanzi2009}.}
\label{NatFig1}
\end{figure*}

In this paper we discuss the possibility to evidence a gravitational contribution, or at least to provide significant bounds, via precision measurements of the cross section near the threshold production of particles and antiparticles. The attention is focused on the cases which appear cleaner from both the experimental and theoretical standpoints. This leads us to consider electron-positron annihilation processes as initial states, producing particle-antiparticle pairs without experiencing color interactions at tree level. The selected final states are therefore the $W^+W^-$, with the advantage of their mass but the drawback of their significantly short lifetime, and the $\mu^+\mu^-$ and $\tau^+\tau^-$, with the advantage of a better statistics, negligible suppression of the cross section enhancement due to the lifetime, also utilizing a comparison with the $e^+e^-$ final state.

The paper is organized as follows. In Sec. II we discuss the modification to the cross section for a particle-antiparticle pair near the production threshold. This is discussed for Yukawa-like potentials, having in mind a contribution of this form arising from a specific parametrization  of strong gravity, and comparing this enhancement to the one of long-range interactions also known as Sommerfeld enhancement.
In Sec. III we discuss the enhancement of the cross section in the most massive case for 
production of colorless particle-antiparticle pairs, the case of $W^+W^-$ production, incorporating its suppression due to the finite lifetime of the $W$s. 
In Sec. IV, using the considerations of the previous two sections, we provide bounds to strong gravity based on measurements of the $W^+W^-$ production performed at LEP-II. 
In Sec. V we provide the complementary example of production of charged heavier leptonic pairs, for which the comparatively longer lifetimes are advantageous, at the price of a reduced expected signal due to the significantly smaller masses, with higher statistics due to the larger cross sections available at those energies. 

These processes, as well as being useful to subtract the  effect of the known Coulomb contribution, could be studied at high precision and high statistics in future electron-positron machines, as qualitatively discussed in the conclusions.

\section{cross section enhancement for particle production with static short-range interactions} 

The creation process of a particle-antiparticle pair is affected in the early stage by the presence of interactions, schematized as static,  and therefore non-relativistic, present between the two particles. For instance, in the simplest case of $e^+e^- \to e^+ e^-$, one expects Coulomb attraction at least. The effect of these potentials is increasingly important as one approaches the production threshold, as the kinetic energy of the produced pair is low, and therefore the two particles are spending more time nearby each other. 

In order to consider a general form of gravitation allowing for a short-distance contribution, we introduce the following potential energy between two pointlike particles of mass $m_1$ and $m_2$
\begin{equation}
V_G(r)=-\frac{G m_1 m_2}{r} \left[1+\alpha_G e^{-r/\lambda_G}\right],
\label{Newton}
\end{equation}
where $\alpha_G$ is a dimensionless coupling constant, expressing the strength of the short-distance component with respect to the known long-distance, Newtonian, component, and $\lambda_G=\hbar/(m_G c)$ is its range, the Compton wavelength of the massive graviton of mass $m_G$ expected from hypothetical strong gravity. The usual long-range Newtonian term is negligible for individual particles, so we will later on consider just the Yukawa term when calculating the strong gravity correction.

The presence of a static potential alters the predictions for the cross sections of annihilation and creation processes, enhancing or suppressing them depending on the attractive or repulsive character of the interaction. In the case of long-range interactions such as the Coulomb one, this effect is named after Sommerfeld \cite{Sommerfeld1931} and in diagrammatic terms corresponds to a ladder diagram in which a virtual particle (photon in the case of Coulomb interactions) is repeatedly exchanged. 
The Sommerfeld (also often named Sommerfeld-Gamow-Sakharov) effect was originally introduced to explain the shift in the energy distribution of the electrons and positrons emitted respectively in $\beta^-$ and $\beta^+$ processes \cite{Wu}. The Coulomb repulsion between the positron and the nucleus after $\beta^+$ emission upshifts its energy, the opposite for electrons after $\beta^-$ emission. The effect is velocity selective, and becomes more relevant at low energy. 

To quantitatively understand the dependence of the enhancement or suppression of the cross section due to a more generic class of potentials upon the relevant quantities, we consider an annihilation process and the effect of the interaction potential on the dynamics of the creation of particles and antiparticles. The interaction potential should in general be treated non-perturbatively, since it will not necessarily be small for small velocities, inducing a modified wave function $\psi$ with respect to the free particle wave function $\psi_0$, both evaluated at the origin (see, for example, \cite{Arkani_Hamed_2009}), $S=|\psi(0)|^2/|\psi_0(0)|^2$. These wave functions are solutions of the Schroedinger equation for a reduced two-body system with and without the static potential, respectively. A finite interaction region in which creation processes do not necessarily occur in the origin leads to a negligible difference in the $S$ factor, as discussed in \cite{Nierop2009}. To evaluate $S$ we need to solve the nonrelativistic Schroedinger equation for scattering problem in the dominant $s$-wave annihilation. The total energy is the kinetic energy associated to the relative motions of the two particles,

\begin{equation}
-\frac{\hbar^2}{M} \frac{d^2 \psi(r)}{dr^2}+V(r)\psi(r)=Mv^2 \psi(r),
\label{Schro}
\end{equation}
with $v$ being their relative velocity.  We use throughout the entire paper units 
in which $\hbar$ and $c$ appear explicitly, unlike natural units.
The solution to Eq. (\ref{Schro}) for the Coulomb interaction yields \cite{Lattanzi2009}

\begin{equation}
S=\frac{\pi \alpha}{\beta} \frac{1}{1-e^{-\pi\alpha/\beta}},
\label{Sformula} 
\end{equation}
where $\alpha$ is the fine structure constant and $\beta=v/c$. 
This relationship shows that in the limit of $\beta \to 0$ the enhancement factor diverges as
 $S\ \sim \beta^{-1}$.

In the case of a generic Yukawa potential  with parameters $\alpha$ and 
$\lambda=\hbar/(m c)$, with $m$ the mass of the intermediate boson

\begin{equation}
V(r) = -\frac{\alpha}{r} e^{-r/\lambda},
\label{Yuk}    
\end{equation}
the Schroedinger equation must be solved numerically. This has been performed 
using the formulation developed in \cite{Iengo2009}, obtaining the dependence upon 
velocity of the particle pair as in Fig. \ref{NatFig1}, for a Coulomb-like potential 
(left plot) and for a Yukawa potential (right plot). 

\begin{figure}[t]
  \begin{center}
  \includegraphics[width=0.98\columnwidth]{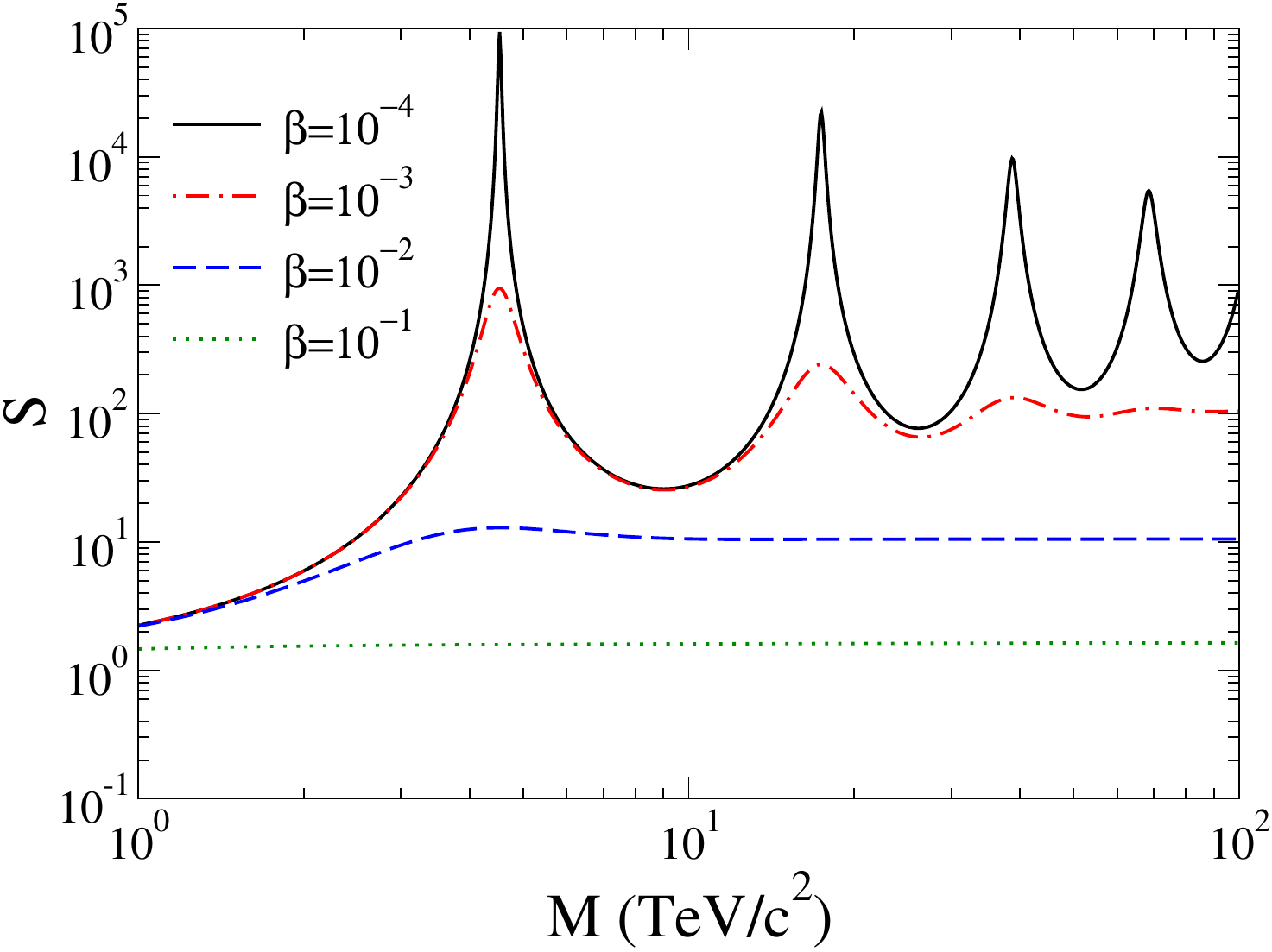}
    \caption{Enhancement of the cross section for a Yukawa potential of coupling strength $\alpha=1/128$ and mass of the intermediate boson $m$= 90 GeV, plotted versus the mass of the created pair of particles $M$, for 
    various values of the particles relative velocity. Resonances appear for masses $M$ above a certain threshold, more pronounced at low velocities.}
    \label{NatFig2}
  \end{center}
\end{figure}

Increasing the mass of the particle pairs leads to an interesting resonant behavior, 
previously discussed \cite{Nierop2009, Iengo2009}, as shown in Fig. \ref{NatFig2}. 
This behavior can be semiqualitatively understood by considering the limit of small distance, 
$r/\lambda <<1$, in which the Yukawa interaction resembles a long-range interaction such as ordinary Newtonian gravitation or the Coulomb interaction.
Indeed, in this limit the exponential in the potential in Eq. (\ref{Yuk}) can be Taylor-expanded to get $V(r)\approx -\alpha/r+\alpha/\lambda$. For small velocities, the potential term dominates over the total energy term, $Mv^2 \ll  \alpha/\lambda$,  and the Schrodinger equation appears as identical to the one for the bound states of a hydrogen atom. We therefore expect resonant peaks at {\it discrete values} of particle masses, for a more detailed derivation of this behavior, see \cite{Lattanzi2009}. 
This feature is retained also when the analysis is numerically extended to the full Yukawa potential, as shown in Fig. \ref{NatFig2}, although with quantitative differences. The resonant behavior is evident especially at low relative velocities, and have been previously discussed in the context of WIMP dark matter detection \cite{Lattanzi2009, Slatyer2010}.

 Note that the mass scale of all known particles is below the resonant regime for an interaction on the order of the weak interaction, so the enhancement increases monotonically with increasing particle mass. Therefore for both types of attractive potential, a larger particle mass leads to a larger enhancement.
This makes the analysis of production of the most massive particles a convenient choice, on top of the natural gain due to 
the presence of the mass in the gravitational potential. 
To avoid complications due to color interaction effects, we focus on $W^+W^-$ production. However, in this case we need to incorporate the corrections due to the finite lifetime of the $W$ boson, which are particularly relevant very near the production threshold, when the time spent by the two 
bosons within their Compton wavelength becomes comparable or longer than their lifetime, as we discuss in the next section.

\section{Production  of unstable particles}

We now consider the annihilation process of the $e^+e^-$ pair into a pair of unstable particles and antiparticles, having in mind the concrete case of the $e^+e^- \to W^+W^-$ process. Indeed, the $W$  bosons have the largest mass for charged particles available so far, considering that top-antitop production is not yet available in electron-positron colliders, therefore they maximize the expected gravitational signal. The choice of colorless final states 
helps to reduce the effect of QCD final state interconnection effects such as 
color reconnection and Bose-Einstein correlations. However, as a consequence of their large mass and therefore of the available phase space for decays, $W$ bosons have a short lifetime. This is affecting the physics around the production threshold, and therefore we expect suppression of the $S$  factor at small relative velocity. The incorporation of the lifetime of final states has been the subject of several studies, starting from \cite{Fadin1988,Fadin1989} in which it has been discussed as an effect protecting the top-antitop production from non-perturbative QCD, potentially leading to toponium bound states, a topic still of current interest. 
The analysis was then extended to the case of W-pair production, including the discussion of higher order effects \cite{Bardin1993a,Bardin1993b,Fadin1993,Fadin1995D,Fadin1995B}.

The general expression for the $W$-pair production cross section from the $e^+e^-$ initial state with off-shell $W$ bosons is given by  \cite{Muta1986}

\begin{eqnarray}
\sigma(s)&=&B_{f_1{\bar{f}_2}} B_{f_3 {\bar{f}_4}} \int_0^s ds_1 \rho(s_1) \times \nonumber \\ 
& & \int_0^{(\sqrt{s}-\sqrt{s_1})^2} ds_2 \rho(s_2) \sigma(s,s_1,s_2),
\label{fullcross}
\end{eqnarray}
where $s_1$ and $s_2$ are the Mandelstam variables for each of the two outcoming $W$ bosons,
$\sigma(s,s_1,s_2)$ the corresponding cross section fully detailed in \cite{Muta1986}, 
$B_{f_1{\bar{f}_2}}$ and $B_{f_3 {\bar{f}_4}}$ are the branching ratios for the decay of the $W$s into fermion pairs 
$f_1{\bar{f}_2}$ and $f_3{\bar{f}_4}$, $\rho(s_i)$ are the Breit-Wigner form factors
\begin{equation}
\rho(s_i)=\frac{1}{\pi} \frac{\sqrt{s_i}~ \Gamma_W(s_i)}{(s_i-M_W^2)^2+M_W^2 \Gamma_W^2(s_i)},
\end{equation}
and $\Gamma_W(s_i)$ is the (running) full width of the $W$ boson, $\Gamma_W(s_i)=3g^2 \sqrt{s_i}/(16\pi)$.
This width accounts for the radiative effects due to the $W$ boson decays \cite{Fadin1995D}, and with respect to \cite{Muta1986} has been updated since the third generation with the top quark does not contribute. It remains to calculate the effect of interactions between the particles on the cross section for $W^+W^-$ $s$-wave production, also including the effect of strong gravity

\begin{eqnarray}
S & =& 1 + 
\frac{\alpha_{C}}{\beta}\delta_{C}+ 
\frac{\alpha_{Z}}{\beta}\delta_{Z}+
\frac{\alpha_{H}}{\beta}\delta_{H}+
\frac{\alpha_{g}}{\beta}\delta_{g} \\
 & = & 1+\sum_{int} \frac{\alpha_{int}}{\beta} \delta_{int},
\label{eq:decoupled}
\end{eqnarray}
where $\beta = 4pc/\sqrt{s}$, and 
\begin{equation}
p=\frac{1}{2c\sqrt{s}}[(s-s_1-s_2)^2-4 s_1 s_2]^{1/2},
\end{equation}
for off-shell $W$ bosons. The terms $\delta_C$, $\delta_Z$ and $\delta_H$ are the expected contributions from standard model interactions, the Sommerfeld one due to photon,  and the ones due to $Z^0$ and Higgs exchange between the W pair, respectively. 
We have also introduced the dimensionless coupling constant $\alpha_g$ 
\begin{equation}
    \alpha_g \equiv {\frac{Gm_1m_2}{\hbar c}}{\alpha_G},
\end{equation}
which allows for a more direct comparison with the other contributions. 
In the case of $W$ bosons ($m_1=m_2=M_W$) we have $\alpha_g = 3.82 \times 10^{-35} \alpha_G$, the largest apart from the one pertaining to a $t\bar{t}$ pair.
Due to the static character of the various potentials,  no interference effects are possible. However, in assessing bounds to strong gravity it is crucial to notice that the largest $\alpha_g$ meaningfully bounded should be smaller than the next-to-leading contribution to $\alpha_C$ in the perturbative expansion of the Coulomb contribution and other sources of signal such as the initial state radiation \cite{{Bardin1993a,Fadin1995B}}, expected to be negligible at the threshold, and amounting to about 0.1 $\%$ in its vicinity \cite{Note}. 

We now evaluate the quantity $\sigma(s, s_1, s_2)$ in Eq. (\ref{fullcross}) defined as $\sigma(s, s_1, s_2) = \sigma_0(s, s_1, s_2) (1+\delta(R, I))$ \cite{Fadin1995D}, where
\begin{equation}
\delta(R,I)=\frac{\alpha_C}{\pi}\delta_{ISR} S \simeq \frac{\alpha_C}{\pi}\delta_{ISR} + 
    \sum_{int} \frac{\alpha_{int}}{\beta}\delta_{int},
    \label{Sum}
\end{equation}
with the Born cross section for off-shell $W$ bosons $\sigma_0(s, s_1, s_2)$ given explicitly in \cite{Muta1986}. In the second line, we use the fact that hard initial state radiation decouples from the static potential contributions to the cross section \cite{Fadin1995D}.  The last relationship in Eq. (\ref{Sum}) holds as long as the corrections to the cross section due to all interactions are perturbative. 

\begin{figure}[t]
\centering
\includegraphics[width=0.98\columnwidth]{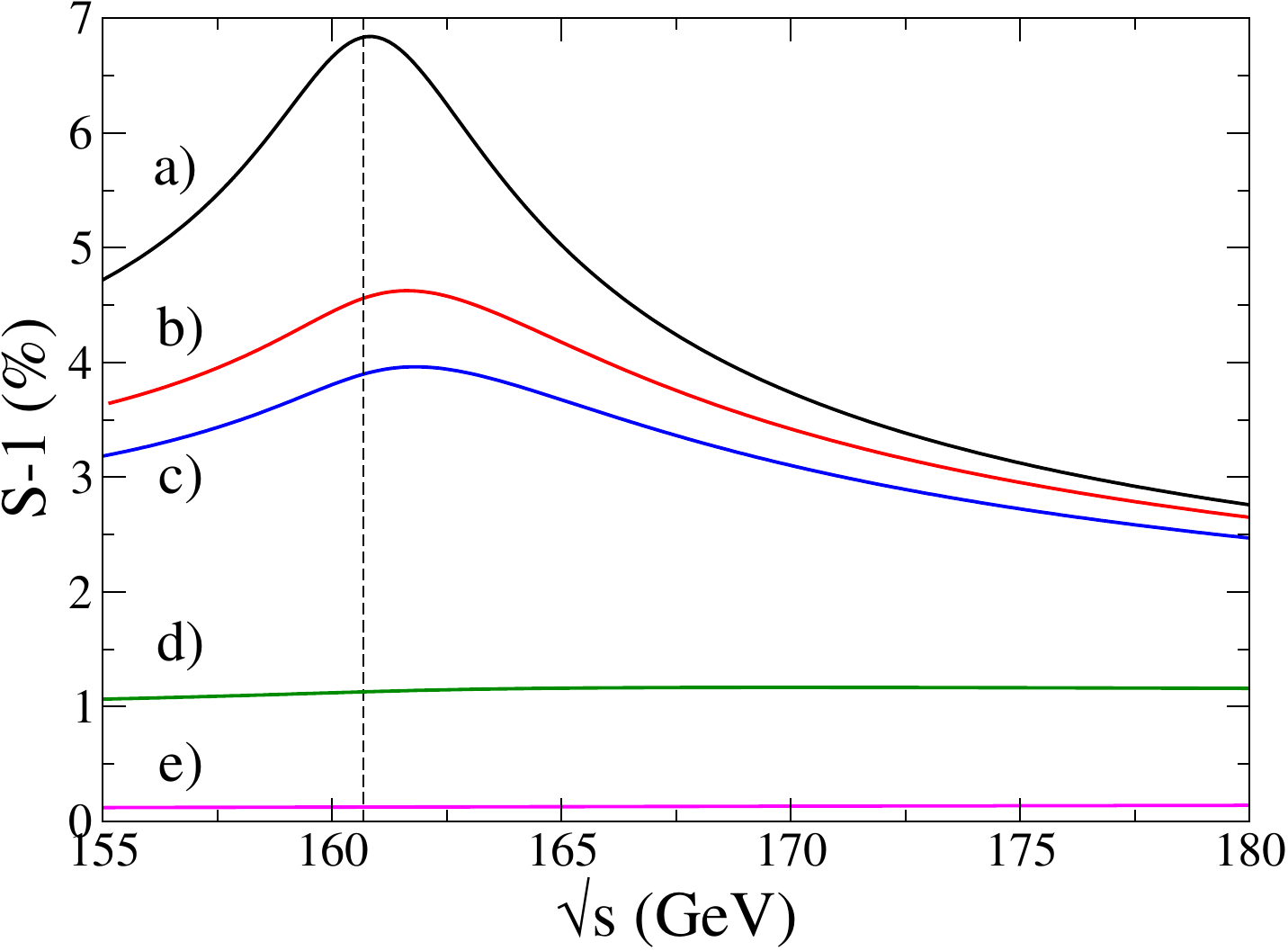}
\caption{Integrated enhancement of the cross section for a short-range gravitational interaction between off-shell $W$ bosons with coupling strength equal to the electromagnetic interaction, versus $\sqrt{s}$. The cases of a massless graviton (therefore contributing identically to a photon) (b), and massive gravitons of mass $m_G$ equal to 10 GeV (c), 100 GeV (d), and 1 TeV (e) are shown, as well as the hypothetical case of stable $W$ bosons (a). The maximum in the enhancement is progressively shifted away from the W-production threshold, and quickly decreasing in amplitude when $m_G$ becomes of the same order of magnitude of $M_W$. The vertical dashed line indicates the on-shell threshold for W-pair production.}
\label{NatFig3}
\end{figure}

For a generic potential $V_{int}(r)$, the corresponding correction is given by \cite{Fadin1995D}

\begin{eqnarray}
\alpha_{int} \delta_{int}&=& -\frac{2}{\hbar} \int_0^{\infty} dr \; V_{int}(r) e^{-p_1 r/\hbar} \times  \nonumber \\
& & \{\sin[(p+p_2)r/\hbar]+\sin[(p-p_2)r/\hbar]\},
\label{Fadin}
\end{eqnarray}
where $p_1$, $p_2$ are defined by 
\begin{eqnarray}
p_1&=&\left[\frac{M_W}{2} \left(\sqrt{E^2+\Gamma_W^2} - E\right)\right]^{1/2},  \\
p_2&=&\left[\frac{M_W}{2} \left(\sqrt{E^2+\Gamma_W^2} +E \right)\right]^{1/2},
\end{eqnarray}
and $E = (s-4M_W^2c^4)/(4M_W c^2)$.

By rewriting the strong gravity potential as 
\begin{align}
    V_g &= -\hbar c \frac{\alpha_g}{r} e^{-r/\lambda_G},
\end{align}
we can solve exactly Eq. (\ref{Fadin}) 
\begin{eqnarray}
& & \alpha_{g} \delta_{g}= -2 \int_0^{\infty} dr \frac{1}{r} e^{-(p_1+\hbar/\lambda_G) r} \times  \nonumber \\
& & \{\sin[(p+p_2)r/\hbar]+\sin[((p-p_2)r/\hbar\} = \\
& & 2\left[\mathrm{arctan}\left(\frac{p+p_2}
    {p_1+m_G c^2}\right)+\mathrm{arctan}\left(\frac{p-p_2}{p_1+m_G c^2} \right) \right],\nonumber
\label{eq:deltaint}
\end{eqnarray}
which also holds for the other Yukawa-like interaction of the $Z_0$ and the Higgs. 
Long-range interactions such as the Coulomb one are just recovered in the $\lambda_G^{-1} \to 0$ limit and appropriate coupling strength \cite{Fadin1995D}.

The enhancement factor for several different Yukawa potentials, all with $\alpha_g=1/128$ and various $m_G$ is plotted as a function of $\sqrt{s}$ in Fig. \ref{NatFig3}. Due to the off-shellness of the $W$ bosons, the enhancement is non negligible at threshold and even below it, without the expected divergent behavior expected at threshold for on-shell particles.  Relative to the case of infinite lifetime (black curve) the finite lifetime suppresses the enhancement, as do values of $m_G$ approaching or larger than $M_W$, thereby  enforcing an effective limitation for searching strong gravity with this method.

\section{Bounds from W-pair production at LEP-II}

We now describe possible bounds to strong gravity from the measurements on $e^+e^- \to W^+W^-$
performed at LEP-II in the $161.3 \;\mathrm{GeV}\leq \sqrt{s} \leq206.6\;\mathrm{GeV}$ range. The lower energy values are 
closer to threshold and will be more affected by all the possible sources of static interaction between the two Ws. 
However the cross section is small in the same region, so statistical errors, for a given integrated luminosity, are large. 
Vice versa, at higher energies the statistical error is smaller, but the expected enhancement is also smaller. 
We then need to assess a range of optimal value of $\sqrt{s}$ for precision measurements, in which one is 
maximizing the sensitivity to the putative strong gravity signal.    

\begin{table}[t]
\begin{center}
\begin{tabular}{| c | c | c | c | c | c | c |}
\hline

$\sqrt{s}$ (GeV) & $\sigma_{\mathrm{exp}}$ (pb) & $S_C$ & $S_Z$ &
$S_H$ & $S_g$ & $\xi$ \\

\hline
161.3 & 3.69  $\pm$ 0.45 & 1.042 & 1.049 & 1.236 & 1.144 & 0.89 \\ 
172.1 & 12.0  $\pm$ 0.7 & 1.029 & 1.051 & 1.253 & 1.115 & 1.48 \\ 
182.7 & 15.92 $\pm$ 0.34 & 1.023 & 1.051 & 1.258 & 1.094 & 3.31 \\ 
188.6 & 16.05 $\pm$ 0.21 & 1.021 & 1.050 & 1.259 & 1.087 & 4.99 \\
191.6 & 16.42 $\pm$ 0.47 & 1.020 & 1.050 & 1.260 & 1.084 & 2.21 \\
\hline
\end{tabular}
\caption{Analysis of the LEP-II measurements on the $e^+ e^- \to W^+ W^-$ channel close to its production threshold. 
In the first column the $\sqrt{s}$ closest to threshold appear, followed by the measured cross section for $W$-pair production, including the error bars, from the analysis of the LEP Electroweak Working Group \cite{LEP200}. The columns from third to sixth contain the $S$ factor for each interaction to which the W pair is sensitive in the Standard Model. The final column contains the values for $\xi \equiv (\bar{S}_{g}-1) \left({\Delta \sigma}/{\sigma}\right)^{-1}$, using the measured $\Delta \sigma$ and $\sigma$ from the second column. The gravitational component of the factor $\bar{S}_g$ is evaluated as described in the text, using the parameters $\alpha_g=1/30$ and $m_G=$ 90 GeV.}
\end{center}
\end{table}

\begin{figure}[t]
  \begin{center}
 {\includegraphics[width=\columnwidth]{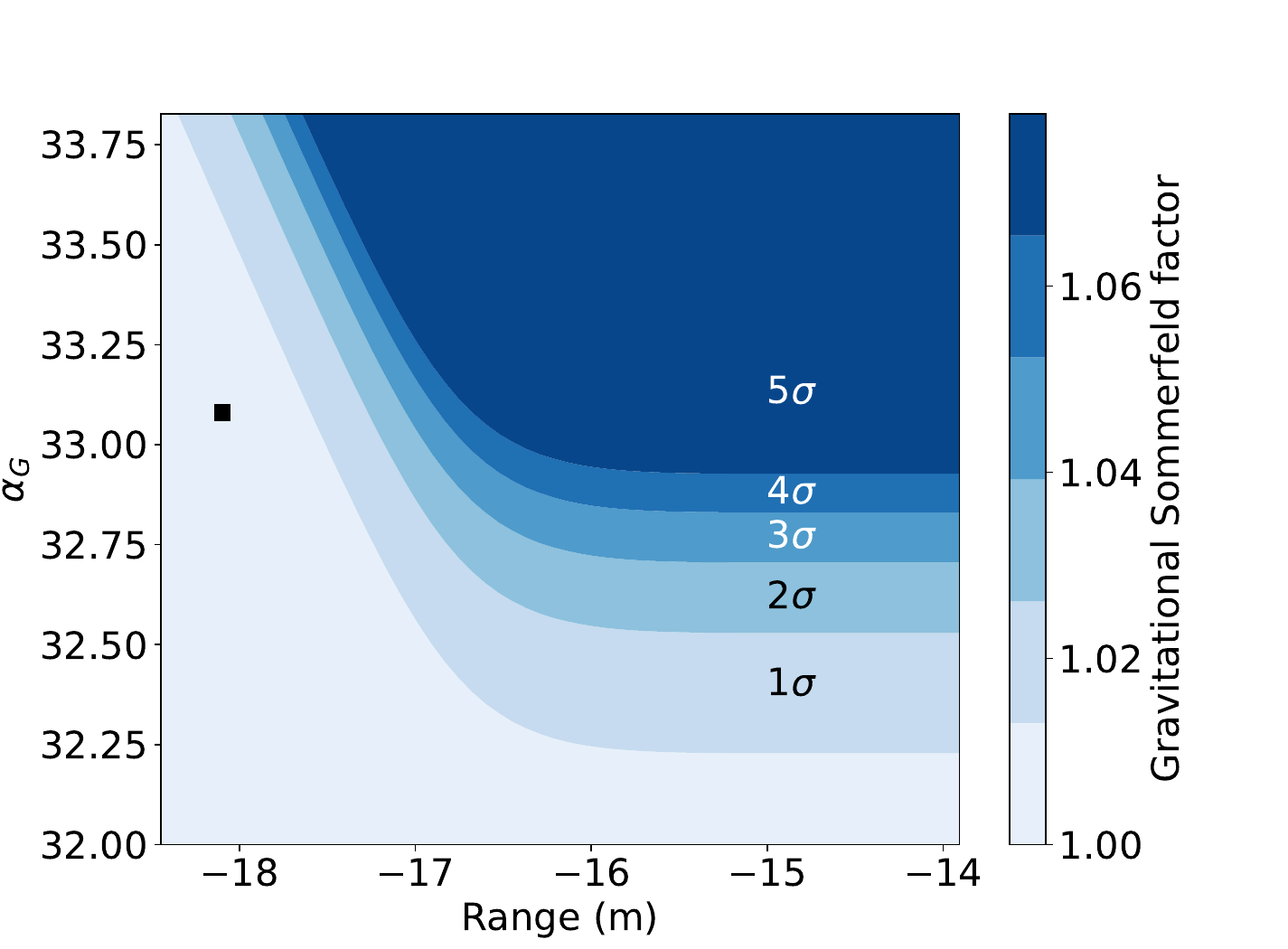}}
    \caption{Contour plot for the exclusion of strong gravity in the $\alpha_G-\lambda_G$ plane, both in log${}_{10}$ scale, from the measured cross section of $e^+e^- \to W^+W^-$ at LEP-II. The black square is the location of the hypothetical gravitoweak unification, at $\lambda_G=8 \times 10^{-19}$ m and $\alpha_G=1.2 \times 10^{33}$, at which gravitation has the same strength as charged weak interactions. The Standard Model cross section used here includes the contributions from photon exchange, $Z^0$ exchange and Higgs boson exchange. We then calculate the ratio of the cross section, including a hypothetical strong gravity interaction, to the Standard Model cross section. Contours delineate the exclusion regions based on the LEP-200 measurements at $\sqrt{s} = 188.6$ GeV. The maximum meaningful value on $\alpha_G$ is imposed by the validity of the perturbation expansion, and corresponds to a dimensionless coupling parameter $\alpha_g \sim 0.25$, therefore $\alpha_G \sim 6.54 \times 10^{33}$, {\it i.e.} 33.82 in the log${}_{10}$ scale.}
    \label{NatFig4}
  \end{center}
  \end{figure}

In terms of measurements of the cross section, we need to take into account the uncertainty on the 
cross section for $W^+W^-$ pair production. To date, there has been no sign of tension between measured and predicted cross section. Therefore the goal is to place bounds on strong gravity parameter space by calculating the $S$ factor for a range of strong gravity potentials, and to then rule out regions of parameter space for which S is too large to agree with observations. Assuming agreement between experiments and the Standard Model prediction, we can obtain confidence contours on $S$ factor using the relative error on the cross section measurement at a given $\sqrt{s}$, the ratio $\Delta \sigma/\sigma$.

We use the cross section measurements taken at LEP-II \cite{LEP200}. The energy measured closest to threshold is $\sqrt{s}=161.3$ GeV. However, we expect to obtain better constraints on strong gravity parameter space using measurements of the cross section taken at higher values of $\sqrt{s}$. After combining the data from the ALEPH, DELPHI, L3, and OPAL collaborations, we collect Table I, containing the center of mass energy, the measured cross section for the production of $W^+W^-$, the contributions to the enhancement due to each exchanged boson expected in the standard model, plus an example of strong gravity contribution to $S$. We also include  the quantity 
 $\xi \equiv (\bar{S}_{g}-1) \left({\Delta \sigma}/{\sigma}\right)^{-1}$, where $\bar{S}_{g}$ is the integrated 
 $S$ factor, given by $\bar{S}_{g} =(\sigma_{SM}+\sigma_g)/\sigma_{SM}$, where $\sigma_{SM}=\sigma_0+\sigma_c+\sigma_z+\sigma_h$, with each quantity being evaluated according to the integration as in Eq. (\ref{fullcross}).
 The statistical bounds are determined by the product value of $\xi$, with a high value of $\xi$ implying greater statistical significance of the measurement to constrain strong gravity parameter space. 

The largest value of $\xi$ in Table I occurs at $\sqrt{s}=188.6$ GeV. We therefore fix $\sqrt{s}$ at this value, and vary $\alpha_G$ and $\lambda_G$ for a constant $\bar{S}_g$ to obtain a plot of the $W^+W^-$ pair production $S$ factor in the strong gravity parameter space (see figure \ref{NatFig4}). 
At this energy, the percent uncertainty on the cross section as measured at LEP is given by $0.21\; {\mathrm{pb}}/16.05\; {\mathrm{pb}} \approx 1.31\%$. We use this relative error to assign statistical significance to the contours in Fig. \ref{NatFig4}, with darker colors corresponding to higher statistical significance in terms of standard deviations. 

As a benchmark, if gravity is embedded into a more extended structure including weak interactions, such that weak interactions are interpreted as its quantum and relativistic  
manifestation at short distances, with the relationship \cite{Onofrio}

\begin{equation}
\frac{G_F}{\sqrt{2}}=\left(\frac{\hbar}{c}\right)^2 \tilde{G}_N,
\end{equation}
where $\tilde{G}_N$ should be a renormalized Newtonian gravitational constant at the attometer scale, we expect this to occur in the $\lambda_G-\alpha_G$ plane at $\mathrm{log}_{10}\alpha_G = 33.08$ and  Yukawa range $8 \times 10^{-19} \mathrm{m}$. This possibility is not yet ruled out, see black square in Fig. \ref{NatFig4}, but it is within an order of magnitude of the 1 $\sigma$ confidence level contour. 

Apart from decreasing the statistical error with larger number of events in future measurements at high luminosity $e^+e^-$ colliders, such as the ones currently under design for the detailed study of the Higgs boson, one possibility to pinpoint a strong gravity contribution consists in studying also the process $e^+e^- \to ZZ$, in which the Coulomb contribution is absent. The data already available on this process from the measurements at LEP-II are not directly competitive with the $W^+W^-$ channel due to the intrinsically lower cross section, generating statistical errors generally larger by one order of magnitude \cite{Jadach,LEP200}.  

\begin{figure}[t]
\begin{center}
{\includegraphics[width=0.98\columnwidth,clip]{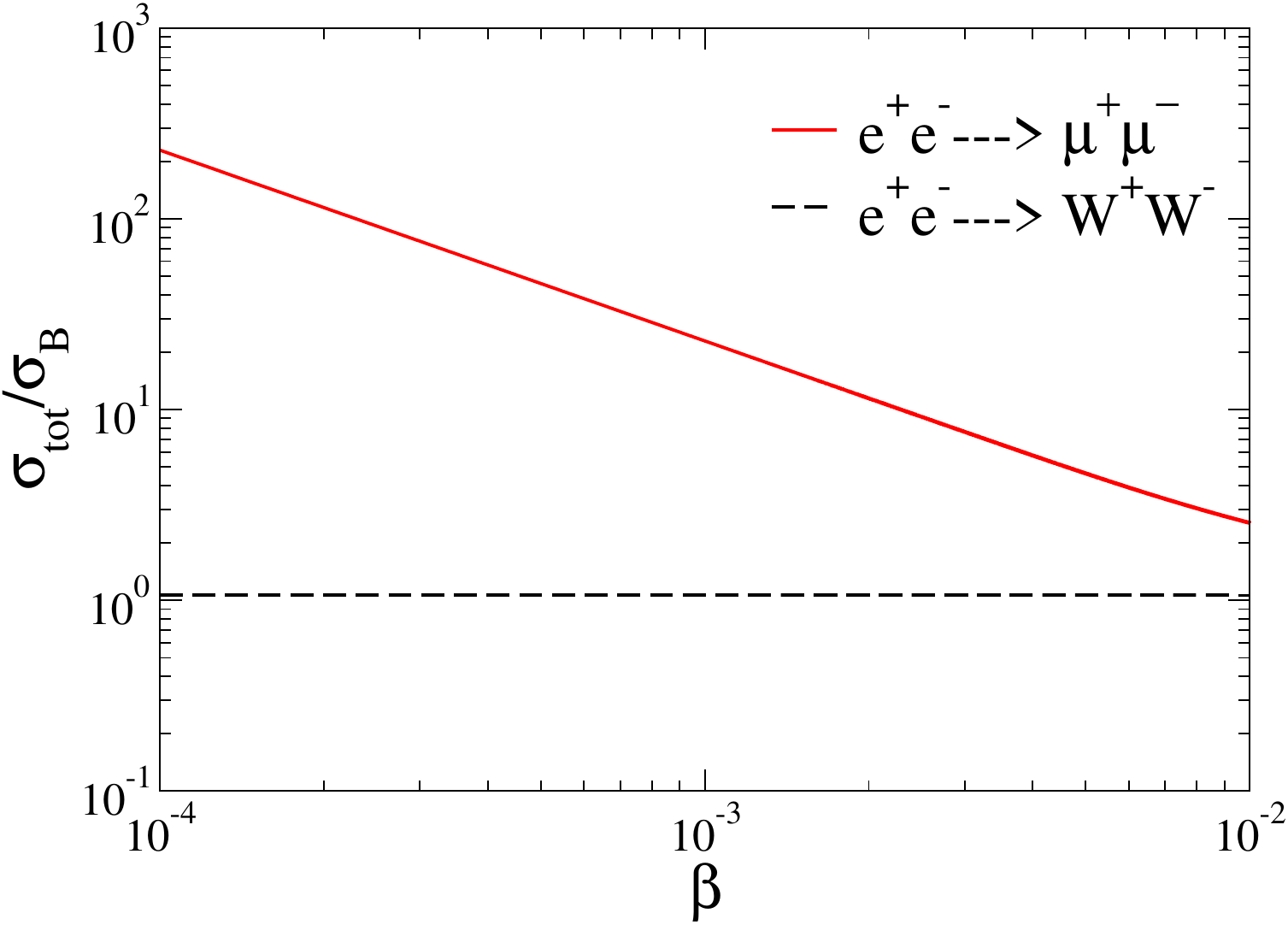}}
    \caption{Ratio between the total cross section for the $e^+ e^- \to \mu^+ \mu^-$ process and the corresponding Born cross section versus $\beta$.
    The same curve also characterizes the $e^+ e^- \to \tau^+ \tau^-$ process, indicating that the lifetime of the final states do not influence the dynamics even at the lowest values of $\beta$ shown. For comparison, the behavior of the same quantity for the $e^+ e^- \to W^+ W^-$ process is shown, with nearly complete suppression of the cross section enhancement.}
    \label{MuonFig}
  \end{center}
\end{figure}

\begin{figure*}[t!]
\centering
\includegraphics[width=\columnwidth,clip]{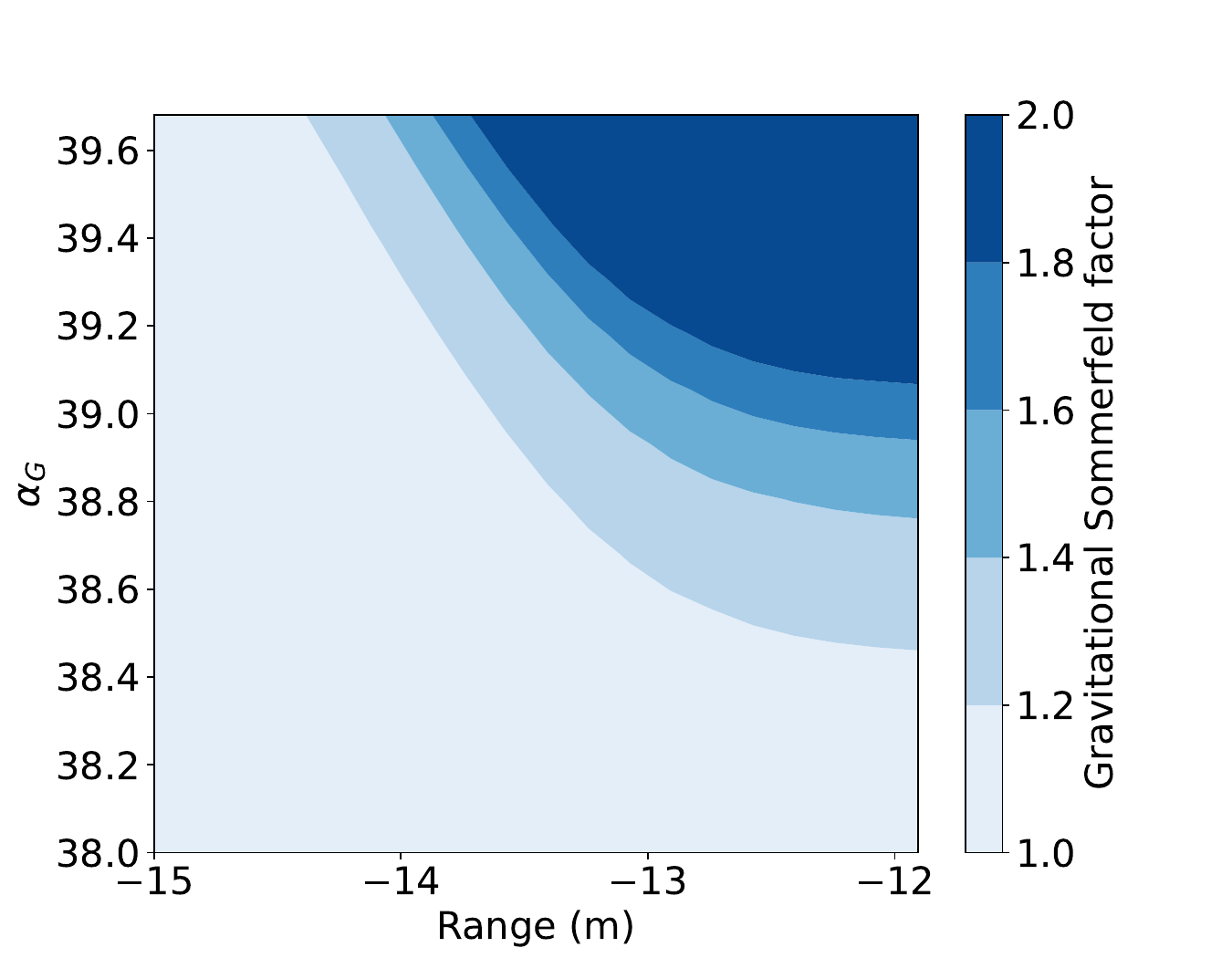}
\includegraphics[width=\columnwidth,clip]{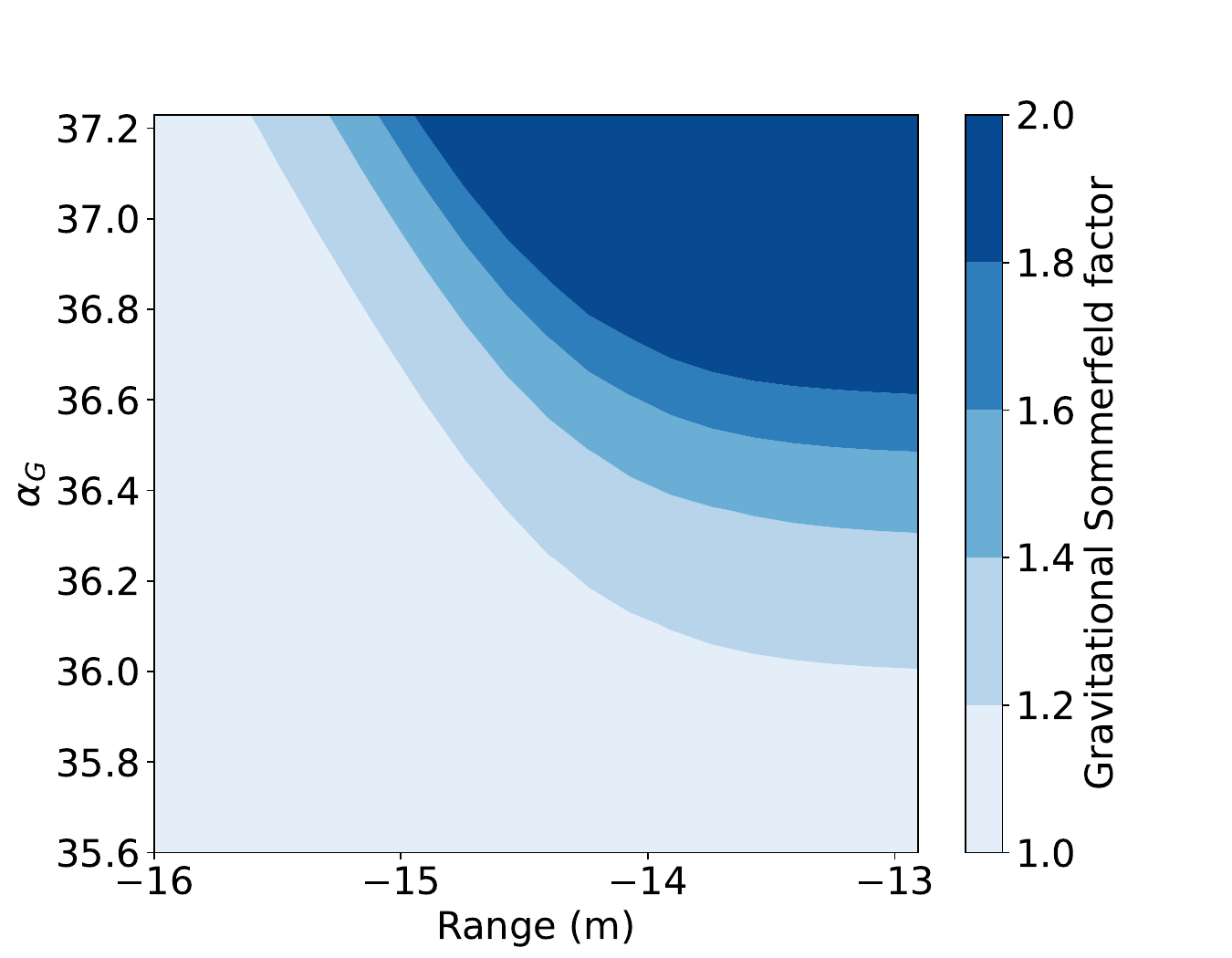}
\caption{Contour plot for the exclusion of strong gravity in the $\alpha_G-\lambda_G$ plane, both in log${}_{10}$ scale, from possible measurements of the $S$ factor in $e^+e^- \to \mu^+ \mu^-$ (left) and in $e^+e^- \to \tau^+ \tau^-$ (right). In both cases the small, but finite width is incorporated. The difference in sensitivity compared to $W^+W^-$ pair production is due to the smaller mass of the final state leptons, which suppresses the gravitational coupling constant $\alpha_g$, with respect to the W pair case, by a factor $5.8 \times 10^5$ for the muon pairs, and $2 \times 10^3$ for the $\tau$ pairs. The false color 
scale on the right side of each plot indicates the magnitude of the expected gravitational 
contribution to the $S$ factor for each projected exclusion region.}
\label{fig:mutauparamspace}
\end{figure*}

\section{Strong gravity in heavy lepton pair production}

This analysis can be naturally extended to the case of $e^+e^- \to \mu^+\mu^-$. Existing electron-positron annihilation experiments have studied the related cross section into muon pairs either with low statistics near the threshold \cite{Borgia,Balakin,Alles}, or more precisely (order of 1$\%$) at higher energies \cite{Akhmetshin,Kurdadze,Ambrosino,Achasov}. Precision measurements of this process are still crucial to better understand the hadronic contribution to the anomalous magnetic moment of the muon. Moreover, if a muon collider is successfully built, it will be possible to achieve similarly near-threshold cross section measurements as LEP. Moreover, the feasibility of muon colliders requires schemes with low transverse emittance beams, and in this context experiments with positron beams slightly above the muon pair production threshold are crucial \cite{Amapane2020}.

The total cross section for $e^+e^- \to \mu^+ \mu^-$ including leading order radiative and initial state radiation corrections has been calculated \cite{Bystritskiy2005}
\begin{equation}
\sigma_{ISR}(s)= \left(1+\frac{2 \alpha}{\pi} \Delta_{ISR}\right) \sigma_B(s), 
\label{sigmaISR}
\end{equation}
where $\alpha$ is the electromagnetic coupling constant $e^2/(\hbar c)$ evaluated at the 
scale $s$, $\sigma_B$ the Born cross section, and  
\begin{eqnarray}
\Delta_{ISR}&=&(\ell_e -1) \times \nonumber \\
& & \left[-\frac{1-3\beta+\beta^3}{\beta(3-\beta^2)} L_\beta-\frac{4}{3} +2 \ln\left(\frac{2\beta}{1+\beta}\right)\right]+ \nonumber \\
& &\frac{3}{4} \ell_e-1+\frac{\pi^2}{6},
\label{Delta}
\end{eqnarray}
where $L_\beta=\ln[(1+\beta)/(1-\beta)]$ and $\ell_e=\ln(s/m_e^2c^4)$. In the limit of small $\beta$ 
Eq. \ref{Delta} becomes

\begin{equation}
\Delta_{ISR}^{(0)}\simeq 
2(\ell_e-1)\left[\ln(2\beta)-1\right]+\frac{3}{4}\ell_e+\frac{\pi^2}{6}-1.
\label{Deltabetazero}
\end{equation}
In this limit, radiative and initial state radiation contributions are quenched, as the static part of the corrections prevails (see \cite{Smith,Voloshin} for a detailed discussion) and therefore we need to interpolate Eq. \ref{sigmaISR} with the following replacement, $\Delta_{ISR}^{\mu^+\mu^-} \to \Delta_{ISR}^{\mu^+\mu^-}-\Delta_{ISR}^{(0)}$, and to incorporate the correction $S$ as a multiplicative factor $1+S$ in Eq. \ref{sigmaISR}.

Taking into account the lifetime of the muon, we plot the enhancement factor with respect to the Born 
cross section, $\sigma_{tot}/\sigma_B$, versus $\beta$ in Fig. \ref{MuonFig}. For comparison, we also show the case of the Ws, in which the lifetime has a significant impact. The long lifetime of the muons of $2.2\mu$s, corresponding to a linewidth $\Gamma_{\mu}=3 \times 10^{-10}$ eV, does not play a role in suppressing the effect of the static potentials, allowing for gains of order $10^2-10^3$ in a range of $\beta$ between $10^{-5}$ and $10^{-4}$. Analogous considerations can be applied to the case of $e^+ e^- \to \tau^+ \tau^-$, with the linewidth of the $\tau$ being $2.27 \times 10^{-4}$ eV, with an enhancement indistinguishable from the one of the muons, but with the benefit of a larger mass. 

We plot a projection of the exclusion region in Fig. \ref{fig:mutauparamspace}, where 
in the absence of actual precision measurements close to threshold we map a possible outcome of the resulting gravitational $S$ factor into the corresponding bounds in the $\alpha_G-\lambda_G$ plane. 
Both these cases do not seem competitive with the bounds available from the W pair production, but a dedicated analysis is in order since the larger cross sections 
at these lower energies can result in smaller statistical errors near the production threshold. In addition, the comparison of the cross section enhancement dependence on $\sqrt{s}$ for the three processes $e^+e^- \to e^+e^-$, $e^+e^- \to \mu^+\mu^-$, 
$e^+e^- \to \tau^+\tau^-$, allows to better disentangle the contributions of the Coulomb interaction from putative strong gravity. For instance, the cross section for the production of lepton pairs increases linearly with $\beta$ near the threshold, and at threshold is inversely proportional to the mass of the produced lepton, $\sigma_0 \simeq \alpha^2 \beta/s$. Due to the inverse dependence of the Coulomb Sommerfeld factor on $\beta$, we expect a finite cross section exactly at threshold 
$\sigma(m_\ell) \simeq \alpha^2/m_\ell^2$, followed by a monotonic
 decrease at higher energies. The strong gravity contribution does not follow the same pattern as the related $S$ factor saturates as discussed in Fig. \ref{NatFig1}. The related cross section is therefore expected to reach a maximum above the threshold and, due to the dependence of the gravitational potential on the square of the mass of the produced leptons, is rather insensitive to the lepton mass. We therefore expect local peaks located at approximately the same distance, in $\sqrt{s}$, from the threshold, regardless of the production of electron, muon and tau pairs. This shows the advantages of a comparative analysis of the various cases. 

\section{Conclusions}

We have shown that measurements of the $W^+W^-$ production cross section near threshold can be used to constrain the parameter space for strong gravity, a sort of microscopic counterpart to the Cavendish experiment to study macroscopic gravitation. We have also briefly discussed the 
potential for analogous bounds using production of heavier leptons close to threshold. The analysis can be refined in the presence of concrete integrated luminosities for various intermediate and high energy electron-positron colliders, using optimization approaches 
as the one discussed in \cite{Stirling}.

More generally, accurate measurements of the cross section can be used to detect Beyond-Standard-Model (BSM) interactions, since any interaction leads to a Sommerfeld-like effect on the 
cross section close to the production threshold. A systematic subtraction of the Coulomb contribution to the Sommerfeld factor can be achieved by measuring the $e^+e^- \to ZZ$ cross section in the case of the $W^+W^-$ channel, and of the 
$e^+e^- \to e^+e^-$ cross section in the case of the heavier lepton production channels.
This can be also relevant to evidence the presence of toponium formation \cite{Fuks}, as well as to understand possible tensions in measurements of the $W$ boson mass \cite{Grzadkowski,Aaltonen2013,Aaltonen2022}.

The bounds on strong gravity presented here are about three orders of magnitude weaker than those presented in \cite{Murata} using LHC data. However, those techniques rely on the presence of a graviton being produced in a process, while the results described above are valid for any model which can be described effectively by a Yukawa potential, for instance in approaches where spacetime itself has a foamy structure at the Planck scale in which gravitons are not necessarily required \cite{Hu}.  Furthermore, the methodology we propose has systematic and statistical errors 
quite different from the ones for direct graviton search, ensuring independent checks in case positive signals are evidenced in the latter scenario. 

These considerations provide further motivation to accurately measure the cross section of fundamental particles near threshold. This would be also feasible either modifying existing machines, such as DA$\Phi$NE and VEPP-2000, or in the first stage of development and calibration of future muon colliders, together with the development of efficient detection of low $\beta$ particles.

\bibliographystyle{apsrev4-1}

\begin{thebibliography}{99}

\bibitem{Dicke} R. H. Dicke, 
Principle of equivalence and the weak interaction,
Rev. Mod. Phys. {\bf 29}, 355 (1957).

\bibitem{Brill} D. R. Brill and J. A. Wheeler, 
Interaction of neutrinos and gravitational fields, 
Rev. Mod. Phys. {\bf 29}, 465 (1957).

\bibitem{Hehl1} F. W. Hehl and B. K. Datta, 
Nonlinear spinor equation and asymmetric connection in general relativity,
J. Math. Phys. {\bf 12}, 1334 (1971).

\bibitem{Hehl2} F. W. Hehl, P. von der Heyde, G. David Kerlick, and J. M. Nester
General relativity with spin and torsion: Foundations and prospects,
Rev. Mod. Phys. {\bf 48}, 393 (1976).

\bibitem{Gasperini}V. De Sabbata and M. Gasperini,
Strong gravity and weak interactions,
Gen. Relativ. Gravit. {\bf 10}, 731 (1979).

\bibitem{Sivaram} V. De Sabbata and C. Sivaram,
{\sl Spin and Torsion in Gravitation} (World Scientific, Singapore, 1994).

\bibitem{Novello} M. Novello and L. M. C. S. Rodriguez, 
A unified model for gravity and electroweak interactions,
Lett. Nuovo Cimento {\bf 43}, 292 (1985).

\bibitem{Loskutov} Yu. M. Loskutov, 
Graviweak interactions and their role in gravitational dynamics and electrodynamics,
J. Exp. Theor. Phys. {\bf 80}, 150 (1995).

\bibitem{Arkani} N. Arkani-Hamed, S. Dimopoulos, and G. R. Dvali, 
The hierarchy problem and new dimensions at a millimeter,
Phys. Lett. B {\bf 429}, 263 (1998).

\bibitem{Antoniadis} I. Antoniadis, N. Arkani-Hamed, S. Dimopoulos, and G. R. Dvali, 
New dimensions at a millimeter to a fermi and superstrings at a TeV,
Phys. Lett. B {\bf 436}, 257 (1998).

\bibitem{Randall} L. Randall and R. Sundrum, 
Large mass hierarchy from a small extra Dimension,
Phys. Rev. Lett. {\bf 83}, 3370 (1999).
  
\bibitem{Dvali} G. Dvali and M. Redi, 
Black hole bound on the number of species and quantum gravity at CERN LHC,
Phys. Rev. D {\bf 77}, 045027 (2008).

\bibitem{Calmet} X. Calmet, S. D. Hsu, and D. Reeb, 
Quantum gravity at a TeV and the renormalization of Newton’s constant,
Phys. Rev. D {\bf 77}, 125015 (2008).

\bibitem{Burinskii1} A. Burinskii,
The Dirac-Kerr-Newman electron,
Gravit. Cosmol. {\bf 14}, 109 (2008).

\bibitem{Burinskii2} A. Burinskii, 
Weakness of gravity as illusion which hides true path to unification of gravity with particle physics,
Int. J. Mod. Phys. D {\bf 26}, 1743022 (2017).

\bibitem{Burinskii3} A. Burinskii,
The Kerr-Newman black hole solution as strong gravity for elementary particles, 
Gravit. Cosmol. {\bf 26}, 87 (2020). 

\bibitem{Dehnen1} H. Dehnen, H. Frommert, and F. Ghaboussi,
Higgs-field gravity,
Int. J. Theor. Phys. {\bf 29}, 537 (1990).

\bibitem{Dehnen2} H. Dehnen, H. Frommert, and F. Ghaboussi,
Higgs field and a new scalar-tensor theory of gravity,
Int. J. Theor. Phys. {\bf 31}, 109 (1992).

\bibitem{Consoli1} M. Consoli,
A weak, attractive, long-range force in Higgs condensates,
Phys. Lett. B {\bf 541}, 307 (2002).

\bibitem{Consoli2} M. Consoli, 
Ultraweak excitations of the quantum vacuum as physical models of gravity,
Classical Quantum Gravity {\bf 26}, 225008 (2009).

\bibitem{Onofrio} R. Onofrio, 
High-energy density implications of a gravitoweak 
unification scenario, 
Mod. Phys. Lett. A {\bf 29}, 1350187 (2014).

\bibitem{Alexander} S. Alexander, A. Marcian\`o, and L. Smolin, 
Gravitational origin of the weak interaction’s chirality,
Phys. Rev. D {\bf 89}, 065017 (2014).

\bibitem{Wheeler} J. A. Wheeler, 
On the nature of quantum geometrodynamics, 
Ann. Phys. 2, 604 (1957).

\bibitem{Abe} K. Abe, {\it et al.} (VENUS Collaboration),
Measurement of the differential cross sections of $e^+e^- \to \gamma \gamma$ and $e^+e^- \to \gamma \gamma \gamma$ at $\sqrt{s}$=55, 56, 56.5 and 57 GeV and search for unstable photino pair production,
Z. Phys. C {\bf 45}, 175 (1989).

\bibitem{Bound1} S. Ask, 
Search for extra dimensions at LEP,
arXiv:hep-ex/0410004v1.

\bibitem{Bound2} ATLAS Collaboration, 
Search for new phenomena with the monojet and missing transverse momentum
signature using the ATLAS detector in
$\sqrt{s}$ = 7 TeV proton–proton collisions,
Phys. Lett. B {\bf 705}, 294 (2011).

\bibitem{Bound3} G. Aad {\it et al.},
Search for dark matter candidates and large extra dimensions in events with a photon and
missing transverse momentum in pp collision data at $\sqrt{s}$=7 TeV with the ATLAS detector, 
Phys. Rev. Lett. {\bf 110}, 011802 (2013).

\bibitem{Murata} J. Murata and S. Tanaka, 
A review of short-range gravity experiments in the LHC era,
Classical Quantum Gravity {\bf 32}, 033001 (2015).

\bibitem{Sommerfeld1931} A. Sommerfeld, 
{\"U}ber die Beugung und Bremsung der Elektronen, 
Ann. Phys. (Berlin) {\bf 3}, 257 (1931). 

\bibitem{Wu} C. S. Wu and R. D. Albert, 
The beta-ray spectra of Cu${}^{64}$,
Phys. Rev. {\bf 75}, 315 (1949).

\bibitem{Arkani_Hamed_2009} N. Arkani-Hamed, D. Finkbeiner, T. Slatyer and N. Weiner, 
A theory of dark matter, Phys. Rev. D {\bf 79}, 1 (2009).

\bibitem{Nierop2009} S. C. A. Nierop, 
The Sommerfeld enhancement, Bachelor thesis, University of Groningen, 2009.

\bibitem{Iengo2009} R. Iengo, 
Sommerfeld enhancement: General results from field theory diagrams, 
J. High Energy Phys. {\bf 05}, 024 (2009).

\bibitem{Lattanzi2009} M. Lattanzi and J. Silk, 
Can the WIMP annihilation boost factor be boosted by the Sommerfeld enhancement?, Phys. Lett. D {\bf 79}, 8 (2009).

\bibitem{Slatyer2010} T. Slatyer, 
The Sommerfeld enhancement for dark matter with an excited state, 
J. Cosmol. Astropart. Phys. {\bf 02}, (2010) 028.

\bibitem{Fadin1988} V. S. Fadin and V. A. Khoze,
Threshold behavior of the cross section for the production of t quarks in $e^+ e^-$ annihilation, 
Pis'ma Zh. Eksp. Teor. Fiz. {\bf 46}, 417 (1987) 
[JETP Lett.  {\bf 46}, 525 (1987)].

\bibitem{Fadin1989} V. S. Fadin and V. A. Khoze,
Production of a pair of heavy quarks in $e^+ e^-$ annihilation in the threshold region,
Yad. Fiz.  {\bf 48}, 487 (1988)
[Sov. J. Nucl. Phys. {\bf 48}, 309 (1988).

\bibitem{Bardin1993a} D. Bardin, M. Bilenky, A. Olchevski, and T. Riemann, 
Off-shell W-pair production in $e^+ e^-$ annihilation. Initial state radiation,
Phys. Lett. B {\bf 308}, 403 (1993).

\bibitem{Bardin1993b} D. Bardin, W. Beenakker, and A. Denner,
The Coulomb singularity in off-shell W-pair production,
Phys. Lett. B {\bf 317}, 213 (1993).

\bibitem{Fadin1993} V. S. Fadin, V. A. Khoze, and A. D. Martin,
On $W^+W^-$ production near threshold,
Phys. Lett. B {\bf 311}, 311 (1993).

\bibitem{Fadin1995D} V. S. Fadin, V. A. Khoze, and A. D. Martin,
Coulomb effects in $W^+W^-$ production,
Phys. Rev. D {\bf 52}, 1377 (1995).

\bibitem{Fadin1995B} V. S. Fadin, V. A. Khoze, A. D. Martin, and W. J. Stirling,
Higher order Coulomb corrections to the threshold $e^+e^- \to W^+W^-$ cross section,
Phys. Lett. B {\bf 363}, 112 (1995).

\bibitem{Muta1986} T. Muta, R. Najima, and S. Wakaizumi, 
Effects of the $W$ boson width in $e^+ e^- \to W^+ W^-$ reactions, 
Mod. Phys. Lett. A {\bf 01}, 203 (1986).

\bibitem{Note} D. Bardin, M. Bilenky, A. Olchevski, and T. Riemann, 
Off-shell W-pair production in $e^+ e^-$ annihilation. Initial state radiation, 
Phys. Lett. B {\bf 357}, 725 (1995), {\bf 308}, (1993) 403(E). 
In table 1 the soft and hard parts of the non-universal QED corrections and the 
second order corrections to the cross section are calculated to be 0.063 pb at 
$\sqrt{s}=190$ GeV, and 0.021 pb at $\sqrt{s}$=176 GeV, respectively 0.38 $\%$ 
and 0.15 $\%$ of the total cross section. 

\bibitem{LEP200} ALEPH Collaboration, Delphi Collaboration, L3 Collaboration, 
OPAL Collaboration, and LEP Electroweak Working Group, 
Electroweak measurements in electron-positron collisions at W-boson-pair energies at LEP, 
Phys. Rep. {\bf 532}, 119 (2013).

\bibitem{Jadach} S. Jadach, W. Placzek, and B. Ward, 
Gauge-invariant YFS exponentiation of (un)stable 
Z-pair production at and beyond CERN LEP 2 energies,
Phys. Rev. D {\bf 56}, 6939 (1997).

\bibitem{Borgia} B. Borgia, F. Ceradini, and M. Conversi, 
Muon pair production by electron-positron collisions in the GeV region,
Lett. Nuovo Cimento {\bf 3}, 115 (1972).

\bibitem{Balakin} V. E. Balakin {\it et al.},
Test of quantum electrodynamics by $e^+ e^- \to \mu^+ \mu^-$,
Phys. Lett. {\bf 37B}, 435 (1971).

\bibitem{Alles} V. Alles-Borelli {\it et al.},
Measurements of $\sigma(e^+e^- \to \mu^{\pm} \mu^{\mp})$ in the energy range 1.2-3.0 GeV,
Phys. Lett. {\bf 59B}, 201 (1975).

\bibitem{Akhmetshin} V. M. Aul'chenko {\it et al.}, 
Measurement of the $e^+e^- \to \pi^+ \pi^-$ cross section with the CMD-2 detector in the 370-520 MeV c.m. energy range,
Pis'ma Zh. Eksp. Teor. Fiz. {\bf 84}, 491 (2006) [JETP Lett. {\bf 84}, 413 (2006)].

\bibitem{Kurdadze} L. M. Kurdadze, M. Lel'chuk, M.; V. Sidorov,  A. Chilingarov, B. Shvarts, 
and S. Eidel'man, 
Study of the decays $\phi \to \mu^+ \mu^-$ and $\phi \to \pi^+ \pi^-$,
Yad. Fiz. {\bf 35}, 352 (1982) [Sov. J. Nucl. Phys. {\bf 35}, 201 (1982)].

\bibitem{Ambrosino} F. Ambrosino {\it et al.}, 
Measurement of the leptonic decay widths of the $\phi$-meson 
with the KLOE detector,
Phys. Lett. B {\bf 608}, 199 (2005).

\bibitem{Achasov} M. N. Achasov {\it et al.}, 
Study of the process $e^+e^- \to \mu^+ \mu^-$ in the energy region 
$\sqrt{s}=080, 1040-1380$ MeV, 
Phys. Rev. D {\bf 79}, 112012 (2009).

\bibitem{Amapane2020} N. Amapane, {\it et al.}, 
Study of muon pair production from positorn annihilation at threshold energy,
J. Instrum. {\bf 15} P01036 (2020).

\bibitem{Bystritskiy2005} Yu. M. Bystritskiy, E. A. Kuraev, G. V. Fedotovich, and F. V. Ignatov,
Cross sections of muon and charged pion pair production in electron-positron annihilation near the threshold, 
Phys. Rev. D {\bf 72}, 114019 (2005).

\bibitem{Smith} B. H. Smith and M. B. Voloshin,
$e^+e^- \to \tau^+ \tau^-$ at the threshold and beyond,
Phys. Lett. B {\bf 324}, 117 (1994).

\bibitem{Voloshin} M. B. Voloshin, 
The onset of $e^+ e^- \to \tau^+ \tau^-$ at threshold revisited,
Phys. Lett. B {\bf 556}, 153 (2003).

\bibitem{Stirling} W. J. Stirling, 
The measurement of $M_W$ from the $W^+W^-$ threshold cross section at LEP2,
Nucl. Phys. B{\bf 456}, 3 (1995).

\bibitem{Fuks} B. Fuks, K. Hagiwara, K. Ma, and Ya-Juan Zheng,
Signature of toponium formation in LHC run 2 data, 
Phys. Rev. D {\bf 104}, 034023 (2021).

\bibitem{Grzadkowski} B. Grzadkowski, Z. Hioki, and J. H. K\"uhn, 
On the $W$ boson mass determination from the total cross section $e^+e^- \to W^+ W^-$,
Phys. Lett. B {\bf 205}, 388 (1988).

\bibitem{Aaltonen2013} T. Aaltonen, {\it et al.} (CDF Collaboration and D0 Collaboration),
Combination of CDF and D0 W-boson mass measurements,
Phys. Rev. D {\bf 88}, 052018 (2013).

\bibitem{Aaltonen2022} T. Aaltonen, {\it et al.} (CDF Collaboration), 
High-precision measurement of the $W$ boson mass with the CDF II detector,
Science {\bf 376}, 170 (2022).

\bibitem{Hu} B. L. Hu, 
Fractal spacetimes in stochastic gravity? - views from anomalous diffusion and the correlation hierarchy, 
J. Phys: Conf. Ser. {\bf 880}, 012004 (2017).

\end{thebibliography}

\end{document}